\documentclass[12pt]{iopart}
\usepackage{epsfig}

\newcommand{\beq}{\begin{equation}}
\newcommand{\eeq}{\end{equation}}
\newcommand{\bea}{\begin{eqnarray}}
\newcommand{\eea}{\end{eqnarray}}
\newcommand{\bce}{\begin{center}}
\newcommand{\ece}{\end{center}}

\def\lsim{\mathrel{\rlap{\lower4pt\hbox{\hskip1pt$\sim$}}
    \raise1pt\hbox{$<$}}}         
\def\gsim{\mathrel{\rlap{\lower4pt\hbox{\hskip1pt$\sim$}}
    \raise1pt\hbox{$>$}}}         

\begin{document}

\title{Theory Highlights of Quark Matter 2004} 

\author{Ralf Rapp\footnote[3]{email: rapp@comp.tamu.edu}  
}

\address{Cyclotron Institute and Physics Department, Texas A\&M University, 
               College Station, Texas 77843-3366, U.S.A.}

\begin{abstract}
Selected highlights of the theoretical developments reported 
at the 2004 Quark Matter conference are discussed, with emphasis 
on open issues.
\end{abstract}




\section{Introduction}
\label{sec_intro}
High-energy heavy-ion collisions (HEHIC's) aim at creating the Quark-Gluon 
Plasma (QGP), together with a thorough
investigation of the phase diagram of strongly interacting matter.
The 17th~International Conference on Ultrarelativistic Nucleus-Nucleus 
Collisions constituted another milestone towards these ambitious 
goals. 

From the theory side, the objective is to develop a coherent and consistent 
description of the relevant processes transforming the incoming nuclei 
into the finally measured (hadronic and electromagnetic) particle 
numbers, spectra, and correlations. The combination of different
approaches is inevitable; e.g., the evolution of bulk matter might
be cast as a succession of color-glass initial state, hydrodynamic
expansion of ultradense phases, and transport simulations of 
moderately dense matter.  To interface these building blocks, knowledge 
on the relevant microscopic degrees of 
freedom, their interactions and spectral properties is mandatory. The 
latter also hold the key to relating observables to (pseudo-) order 
parameters of the phase transition(s) to be identified. A further
challenge is posed by the transition regime from bulk matter 
to perturbative spectra, as emphasized in Wiedemann's theoretical 
overview talk~\cite{Wied04}. 

Along these lines, I review below progress that has been
reported on theoretical approaches to the QCD phase diagram 
(Sec.~\ref{sec_theo}), on properties of bulk matter as deduced from
RHIC runs 2+3 (Sec.~\ref{sec_bulk}) and on microscopic probes 
expected to be scru\-tinized in run 4 and the recent SPS run 
(Sec.~\ref{sec_probes}).  Final remarks are made in Sec.~\ref{sec_sum}.   

\section{QCD Theory at Finite Density and Temperature}
\label{sec_theo}

\subsection{Colorsuperconductivity}
\label{ssec_csc}
The richness of the phase diagram at high densities and {\em small}
temperatures continues to provide for new phases, especially under
conditions relevant for compact stars such as color and charge 
neutrality. The key quantity at moderate quark chemical potential, 
$\mu_q$$\simeq$~400~MeV, is the (constituent) strange quark mass, 
$m_s^{(*)}$. Rajagopal elucidated that, reducing $\mu_q$ from large 
values, the first transition is probably from the $u$-$d$-$s$ symmetric 
CFL ground state into a ``gapless" CFL phase~\cite{AKR03}: the 
asymmetry induced by $m_s$ trifurcates the gap value and induces 
``blocking" regions on the $d$- and $s$- Fermi surfaces where their 
pairing is inhibited. One of the open questions is whether a 2SC state 
($u$-$d$ pairing only) exists in nature, which essentially hinges on 
the condition $\mu_q < m_s^*$ being met in the quark phase, thus 
depending on nonperturbative forces responsible for chiral 
symmetry breaking ($\chi$SB) in the vacuum.    

Applications to neutron star phenomenology were discussed by 
Reddy, who argued that star radii of 8~km or less clearly favor
a quark-matter equation of state (EoS). Much like in HEHIC's, 
electroweak emission is a promising signal of in-medium effects, 
probing the low-lying excitations (Goldstone bosons,  
(di)quark quasiparticles) of the matter in, e.g., 
supernovae explosions or long-term cooling~\cite{RST03,JPS02,VRO03}.

\subsection{Chiral vs. Deconfinement Transition}
\label{ssec_trans}
\begin{figure}[!t]
\begin{center}
\epsfig{file=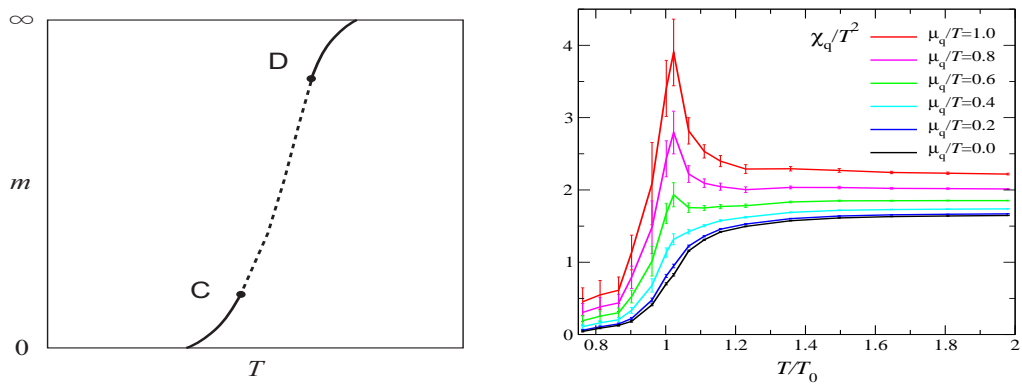,width=6.1cm,height=5cm}
\hspace{1cm}
\epsfig{file=chiqT_0424C.eps,width=6.1cm,height=5cm}
\end{center}
\vspace{-0.4cm}
\caption{Left panel: first order lines (solid) and critical endpoints
for chiral (lower, C) and deconfinement (upper, D)
transitions in the $T$-$m_q$ plane for 2+1 flavor QCD~\cite{Gav94,HF03}.
Right panel: quark number susceptibilities for different chemical
potentials~\cite{All03}. }
\label{fig_lat}
\end{figure}

A longstanding problem associated with the QCD phase diagram is the 
relation between chiral and deconfinement transitions. Current lattice 
calculations indicate that in real QCD the small but finite current 
quark masses, $m_q$, lead to a finite-$T$ transition which is not of 
first order. However, the peaks in the susceptibilities pertinent to 
order para\-meters at $m_q$$\to$$\infty$ and $m_q$=0 -- Polyakov loop 
$\langle L \rangle$ and quark condensate $\langle \bar qq\rangle$, 
respectively -- precisely coincide~\cite{Karsch02}.  
Fukushima suggested~\cite{HF03} this feature to arise through a mixing
of the scalar ``soft" modes that become massless at the second order 
endpoints in the $T$-$m_q$ plane, cf.~left panel of Fig.~\ref{fig_lat}.
For the chiral transition this is the usual ``$\sigma$"-meson, 
whereas for deconfinement he argued it to be the electric
glueball field. Their mixing ensures the coincidence of the (pseudo-)
transitions also in the cross-over region. It is not quite clear, 
however, why this mechanism is not operative for quarks in the adjoint 
representation where lattice QCD (LQCD) observes two distinct transition 
temperatures~\cite{KL99}. 
Similar issues were addressed by M\'ocsy~\cite{MST03}.

\subsection{Lattice QCD Results}
\label{ssec_lat}
One can close the gap between the two endpoints $C$ and $D$ 
in Fig.~\ref{fig_lat} (left panel) by moving to finite $\mu_q$. 
Recent lattice computations have pursued this by an expansion in 
$\mu_q/T$, as discussed by Redlich and Karsch~\cite{All03,KRT03}. 
Fig.~\ref{fig_lat} (right panel) shows
results for the quark number susceptibility, which indeed develops
a rather sharp maximum (signaling a rapid change in 
baryon density) at $\mu_B$=$3\mu_q$$\simeq$500~MeV, quite reminiscent 
to independent deter\-minations of the critical endpoint~\cite{FK04}. 
This puts the latter into the realm
of the future GSI facility, the physics of which was reviewed 
by Friman~\cite{Fri04}.    

Further progress on finite-$T$ (quenched) LQCD results for 
charmonium properties was reported by Karsch, Asakawa and Datta.
From the color-singlet free energy, $F_1=V_1-TS$, the pertinent
$c\bar c$ potential has been extracted which appears to support
bound states for a range of temperatures above $T_c$~\cite{Kacz03}. 
This corroborates the analyses of spectral functions exhibiting 
resonance/bound states of low-lying charmonia ($\eta_c$, $J/\psi$) 
surviving up to $\sim$2$T_c$~\cite{AH04,Datt03}, which could have 
important consequences for charmonium production at RHIC, see 
Sec.~\ref{ssec_charm} below.   

\section{QCD Bulk Properties at RHIC}
\label{sec_bulk}
In this section, I will essentially follow a chronological order in 
discussing the properties of matter as evolving in a HEHIC, i.e., from 
the initial state (Sec.~\ref{ssec_cgc}) via the 
dense phases through the phase boundary 
(Secs.~\ref{ssec_ot},~\ref{ssec_hc}) until thermal freezeout 
(Sec.~\ref{ssec_hbt}).

\subsection{Color-Glass Condensate (CGC)}
\label{ssec_cgc}
\begin{figure}[!t]
\begin{center}
\epsfig{file=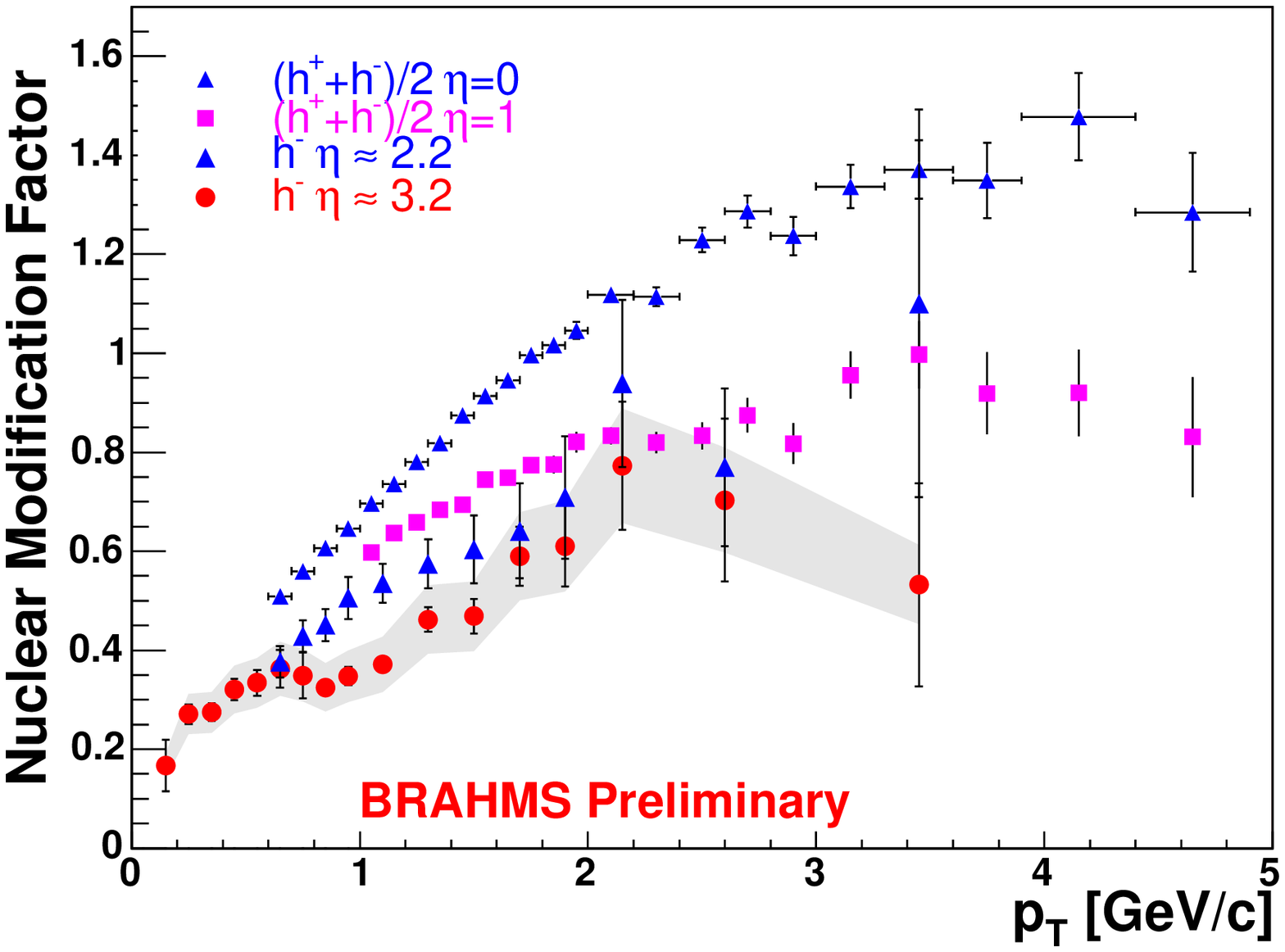,width=6.6cm,height=6.2cm}
\hspace{0.5cm}
\epsfig{file=jet-corr.eps,width=6.1cm,height=6.4cm}
\end{center}
\vspace{-0.4cm}
\caption{Left panel: rapidity dependence of $R_{dAu}(p_t)$ in 200~GeV
$d$-$Au$~\cite{Deb04}. Right panel: STAR data~\cite{star-dAu} and 
jet quenching calculations~\cite{Vit03} for azimuthal correlations.}
\label{fig_dA}
\end{figure}

An important experimental finding since the previous Quark Matter
conference is the absence of a suppression of (moderately)
high transverse-momentum ($p_t$) hadrons in 200~GeV $d$-$Au$ collisions 
at midrapidity ($\eta$=0), which, however, turns into a suppression at 
forward $\eta$~\cite{Deb04}, cf.~left panel of Fig.~\ref{fig_dA}. 
Whereas the former confirmed strong suppression of high-$p_t$ hadrons
in $Au$-$Au$ as an energy loss effect in the produced hot/dense medium
(see next section), the latter might have provided first explicit 
evidence of gluon saturation in the low-$x$ part of the $Au$ nucleus'
wave function, as advocated by Jalilian-Marian, 
Venugopalan~\cite{JNV03} and Kovchegov (see also Ref.~\cite{McLe04}). 
In addition, the increase of suppression with 
centrality of the $d$-$Au$ collision is qualitatively in line with the 
CGC~\cite{KKT02}. Definite conclusions have to await quantitative 
comparison with data incorporating also more mundane mechanisms,  
as stressed by Accardi~\cite{Acc04}.
E.g., the HIJING event generator~\cite{hijing} satisfactorily describes 
the rapidity spectra in $d$-$Au$~\cite{brahms03}, as well as the 
$h^+/h^-$ asymmetry in the forward region, indicative for valence-quark 
fragmentation.  The former implies that the low-$p_t$ regime is not 
an unambiguous signal of CGC. The critical $p_t$-region 
will thus be around the estimated saturation momentum 
of $Q_s$$\simeq$2~GeV (for $\eta$=3 at RHIC). Forward charge asymmetries
further complicate the interpretation of the nuclear modification
factor $R_{dAu}$ which uses a $p$-$p$ reference as denominator.
Rapidity energy loss of projectile valence quarks can be
expected to induce a decreasing centrality dependence of $R_{dAu}$,
too.

One should note that if the CGC is indeed operative at moderate 
$p_t$ in forward $d$-$Au$,
it will reflect itself also in bulk hadron production systematics
in $Au$-$Au$ at midrapidity. Using pertinent initial conditions
for a hydrodynamic evolution leads to a satisfactory 
description of the observed $dN/dy$ distributions, as shown in
Nara's talk~\cite{Nara04}.  

\subsection{Opacity and Thermalization}
\label{ssec_ot}
The main evidence for the production of thermalized matter at RHIC,
at energy densities well above the critical one, resides on the 
quantitative success of parton energy-loss and hydrodynamic calculations. 

As discussed by Vitev and Barnaf\"oldi, using well calibrated 
perturbative QCD probes in connection with mostly medium-induced
(non-abelian) gluon radiation allows to infer energy densities
$\epsilon$$\simeq$20~GeV/fm$^3$ in the early stages of 200~AGeV central 
$Au$-$Au$ collisions. In addition to correctly predicting 
the factor of 4-5 reduction of the nuclear modification factor, 
$R_{AA}(p_t)$, together with its flatness up to the highest currently 
measured $p_t$$\simeq$11~GeV, the calculations reproduce azimuthal 
correlations, especially for the away-side jet, such as its (slight) 
broadening in $d$-$Au$~\cite{Vit03}, its gradual disappearance with 
centrality in $Au$-$Au$~\cite{Wang03} (cf.~right panel of 
Fig.~\ref{fig_dA}), as well as its preferably in-plane reappearance 
in semi-central $Au$-$Au$ (in the short direction of the ``almond"). 

The relevance of ``hadronic" quenching was addressed by Greiner.
Whereas Lorentz-dilation precludes noticeable impact on formed
hadrons, absorption of colorless ``pre\-hadrons" (emerging from 
hard-scattered valence quarks color-neutralizing on time\-scales 
$t$$\sim$$E_{jet}/Q^2$$\sim$$1/p_t$~\cite{Kop03}) on surrounding
(pre-) hadrons may be significant. However, the implementation of 
this approach in a transport model~\cite{Cass03} tends to 
over\-estimate high-$p_t$ suppression at SPS, while underestimating 
elliptic flow at RHIC.  

Ideal hydrodynamics is remarkably successful in describing the 
collective evolution of $\sim$ 99\% of all produced particles 
at RHIC as illustrated in Hirano's talk (corrections away from 
midrapidity and due to viscosity were discussed by 
Heinz and Teaney, respectively).
The inherent early thermalization time, $\tau_0$$\simeq$0.5~fm/c,
required especially for second and fourth order azimuthal asymmetries,  
$v_2(p_t)$ and $v_4(p_t)$, however, cannot be accounted for 
in terms of perturbative partonic rescattering cross sections, 
which are much too small.
\begin{figure}[!t]
\begin{center}
\epsfig{file=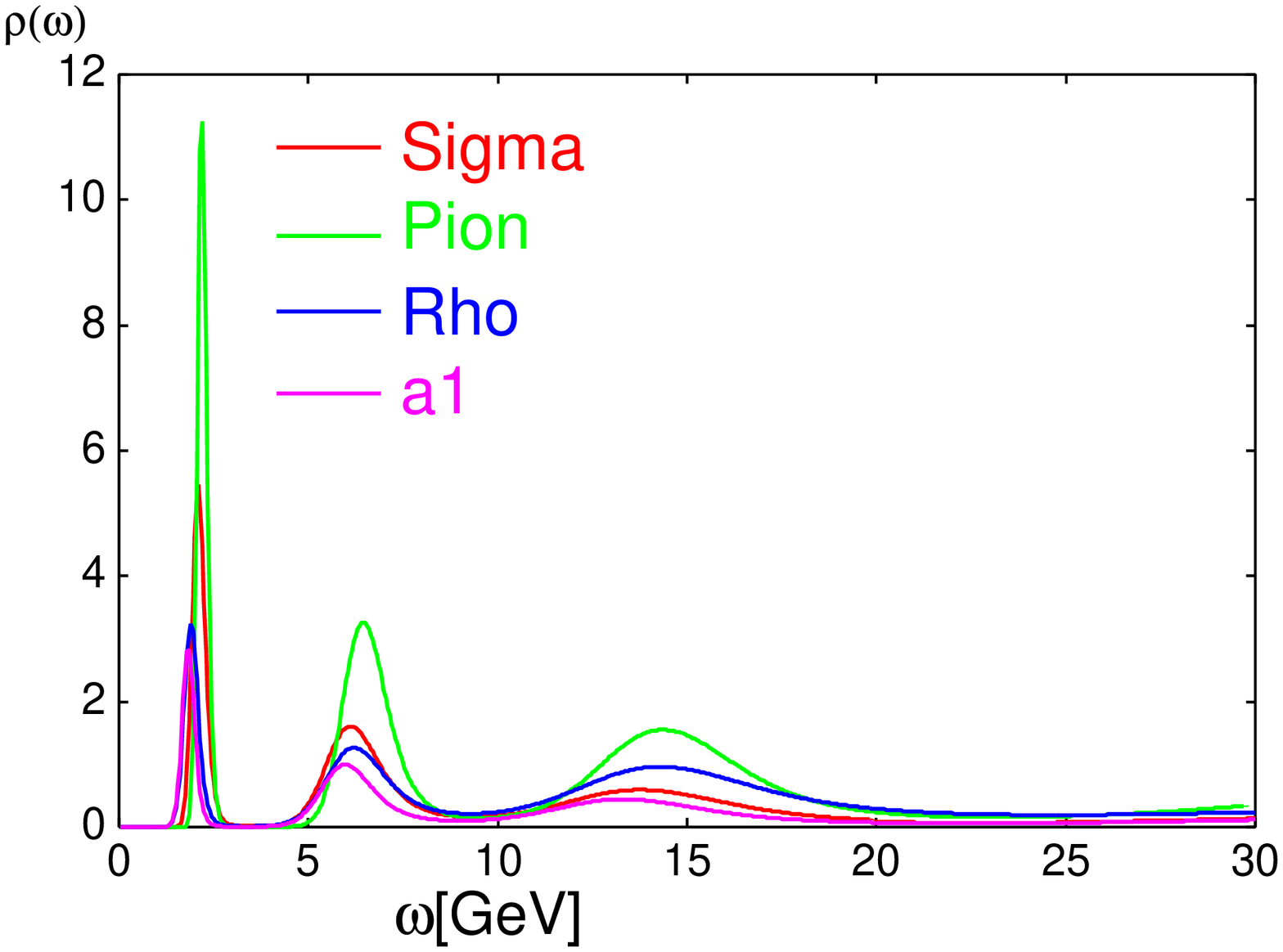,width=6.2cm,height=5.5cm}
\hspace{1cm}
\epsfig{file=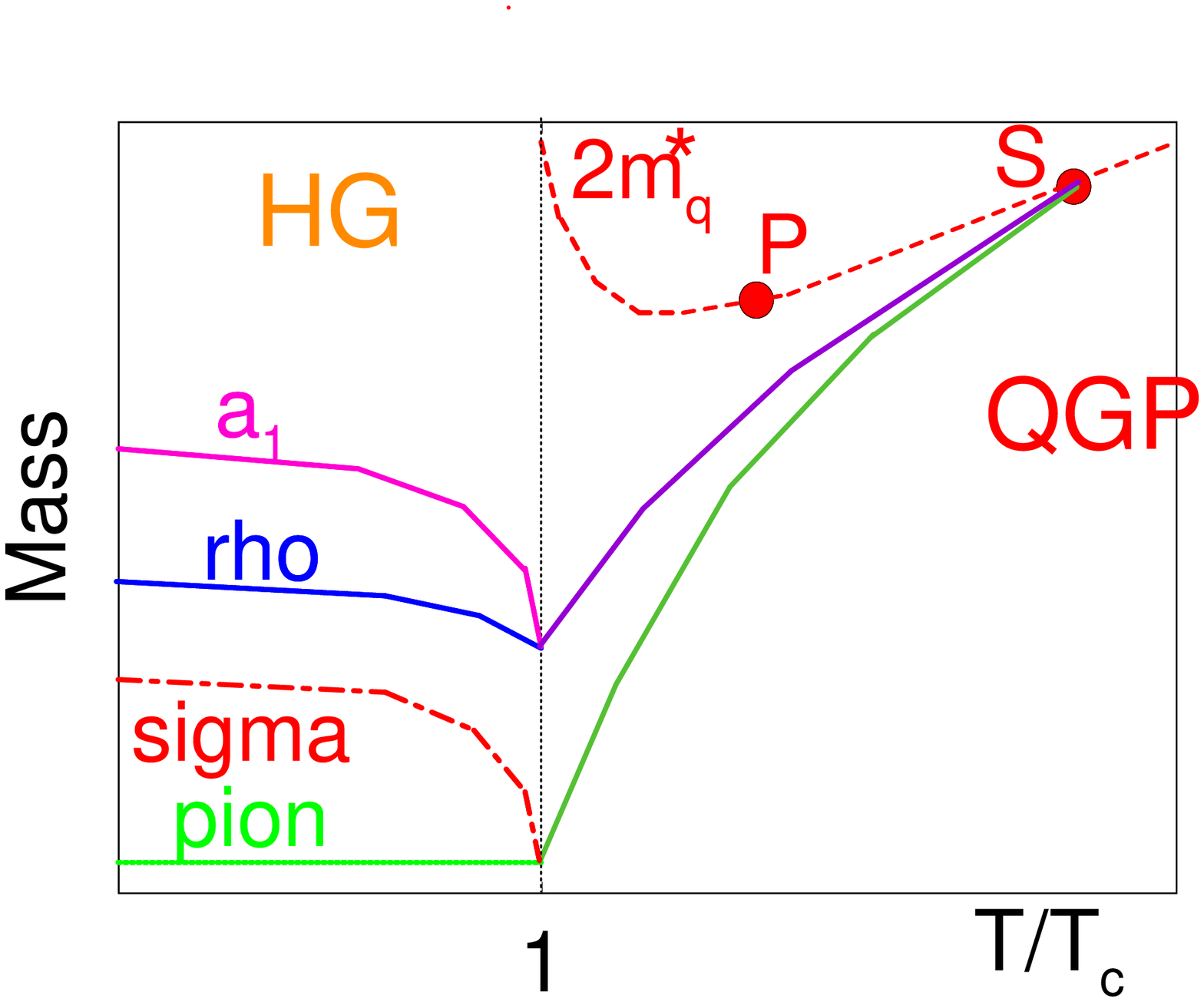,width=6.0cm,height=5.5cm}
\end{center}
\vspace{-0.4cm}
\caption{Left panel: spectral functions for light hadrons
as extracted from quenched LQCD calculations~\cite{AH03}.
Right panel: schematic dependence of light hadrons masses
across the phase boundary, with $S$- and $P$-wave dissolution
points in the QGP~\cite{BLRS03}.}
\label{fig_qgp}
\end{figure}
Here, progress could lie in the realization that for moderate 
temperatures (1-3~$T_c$) the QGP may support $q$-$\bar q$ and/or 
$g$-$g$ bound states (resonances), as evidenced by recent LQCD 
calculations of ``hadronic" spectral functions~\cite{AH03}, cf.~left 
panel of Fig.~\ref{fig_qgp}. Shuryak and Zahed~\cite{SZ03} suggested 
the color-Coulomb interaction as the underlying force, with the 
running of $\alpha_s$ only limited at the screening scale $m_{Debye}$.  
Brown argued that, close to $T_c$, additional instanton-mole\-cule ($IM$) 
induced interactions (augmented by RPA resummations and spatially small
Cou\-lomb wave functions) become important to bind thermal quasi-quarks 
into (almost) massless pions~\cite{BLRS03}, being continuously 
connected to the hadronic phase (even though standard estimates of 
$IM$ coupling strengths are rather small~\cite{SSV95}),
cf.~right panel of Fig.~\ref{fig_qgp}.    

Resonant rescattering of partons is a promising mechanism to explain 
the short thermalization times required by hydrodynamics, provided
the resonance correlations themselves build up sufficiently fast.


\subsection{Hydrodynamics vs. Quark Coalescence}
\label{ssec_hc}
Relativistic hydrodynamics, employing an equation of state consistent
with predictions from LQCD, provides an excellent description
of low-$p_t$ hadron spectra. Imple\-menting jets and their energy loss
into a hydrodynamic evolution, Hirano and Nara confirmed that the 
transition momentum from collective to perturbative production
increases with mass (2, 2.6 and 3.5~GeV for $\pi$, $K$ and $p$ in 
central $Au$-$Au$, respec\-tively)~\cite{HN03}. However, even the hydro 
protons appear to ``run out of steam" too early, in that the $p$/$\pi$ 
ratio in the hydro+jet model underpredicts the experimental 
data~\cite{phenix-ppi} above $p_t$$\simeq$3~GeV (similarly, the 
experimental $\Lambda/K$ ratio recovers the perturbative value only 
close to $p_t$$\simeq$6~GeV~\cite{star-laka}). Furthermore, the elliptic 
flow of all thus far identified hadrons appears to follow a 
constituent-quark scaling, i.e., a universal ``partonic"  
$v_2(p_t/n)/n$ with $n$=2 (3) for mesons (baryons). Both the
universal $v_2$ and the enhanced baryon-to-meson ratios at intermediate
$p_t$ are readily explained within a quark coalescence picture
at hadronization, as shown in the talks by Fries, Hwa and Molnar.  
Whereas the model of Fries~{\it et~al.}~\cite{Fries03} 
focuses on coalescence of thermal partons, 
Greco~{\it et~al.}~\cite{Greco03} and  
Hwa~{\it et~al.}~\cite{Hwa03} also allow for recombination with 
minijet partons.  Whether the latter will be able to explain the 
observed nearside correlations in the 4~GeV regime (recall right panel 
of Fig.~\ref{fig_dA}) -- which appear unmodified from $p$-$p$ to 
central $Au$-$Au$ -- remains an open question at present. In fact, 
in the approach of Hwa~{\it et~al.}~\cite{Hwa03}, also
in $p$-$p$ collisions a significant contribution is assigned to
coalescence.  
As for the distinction from hydrodynamic models, the elliptic flow of 
$\phi$ mesons (with a mass similar to baryons but only $n$=2 
constituent quarks) is an ideal observable. First data on the ratio 
of central to peripheral $\phi$ $p_t$-spectra, $R_{CP}^\phi(p_t)$, 
seem to follow meson systematics.\footnote{One should note
that hydrodynamics ceases to be applicable in peripheral collisions.}

In view of the LQCD spectral functions (Fig.~\ref{fig_qgp}), a logical 
extension of coalescence models could include hadron formation even 
above $T_c$, e.g., by solving suitable rate equa\-tions, reminiscent to 
recent calculations in the charmonium context~\cite{GRB03} 
(see Sec.~\ref{ssec_charm}).

\subsection{HBT ``Puzzle"}
\label{ssec_hbt}
New suggestions have been presented at this meeting for the problem 
of the HBT data, especially the smallness of the ``out"-radius,
$R_{out}^2=D(x_{out},x_{out})-2D (x_{out}, \beta_t t)+D(\beta_t t,
\beta t)$  ($D$: variance). 
Kapusta~\cite{Kap04} and Wong~\cite{Wong04} argued that 
keeping track of quantum phases in the hadronic rescattering could 
preserve memory on the initial source size, thereby also addressing the 
approximate constancy of the radii with collision energy. 

We recall that the AMPT transport model~\cite{LKP02} can 
account for the measured radii, with a positive $x_{out}$-$t$ 
correlation, (partially) related to decays of long-lived resonances 
(especially $\omega$'s), as a key ingredient to reduce $R_{out}$. 
The question re\-mains where the discrepancy to other 
transport models~\cite{Soff03} resides, which should be rather 
similar in the treatment of the
late stages, thus pointing to differences in earlier stages.

\section{Microscopic Probes of QCD Matter}
\label{sec_probes}
With the bulk properties of the produced matter at RHIC being 
reasonably well assessed, the next challenge is to determine its
microscopic properties. In this section, I will focus on 3 complexes 
of observables that are expected to serve this goal, 
proceeding from the relatively dilute freezeout to the hot and dense
phases.

\subsection{Resonance Spectroscopy}
\label{ssec_res}
Short-lived resonances are a promising tool to deduce in-medium 
modifications of their spectral shape close to thermal freezeout 
through invariant-mass spectra of their decay products, 
e.g., $\rho^0\to\pi^+\pi^-$ or $\Delta^{++}\to p\pi^+$~\cite{star-res}. 
For $(\mu_N,\mu_\pi,T)_{fo}$=(370,90,110)~MeV, the 
pion (anti-/baryon) density is, in fact, still appreciable, 
$\varrho_\pi$=0.65$\varrho_0$ ($\varrho_{B+\bar B}$=0.25$\varrho_0$),
with $\varrho_0$=0.16~fm$^{-3}$.
\begin{figure}[!t]
\begin{center}
\epsfig{file=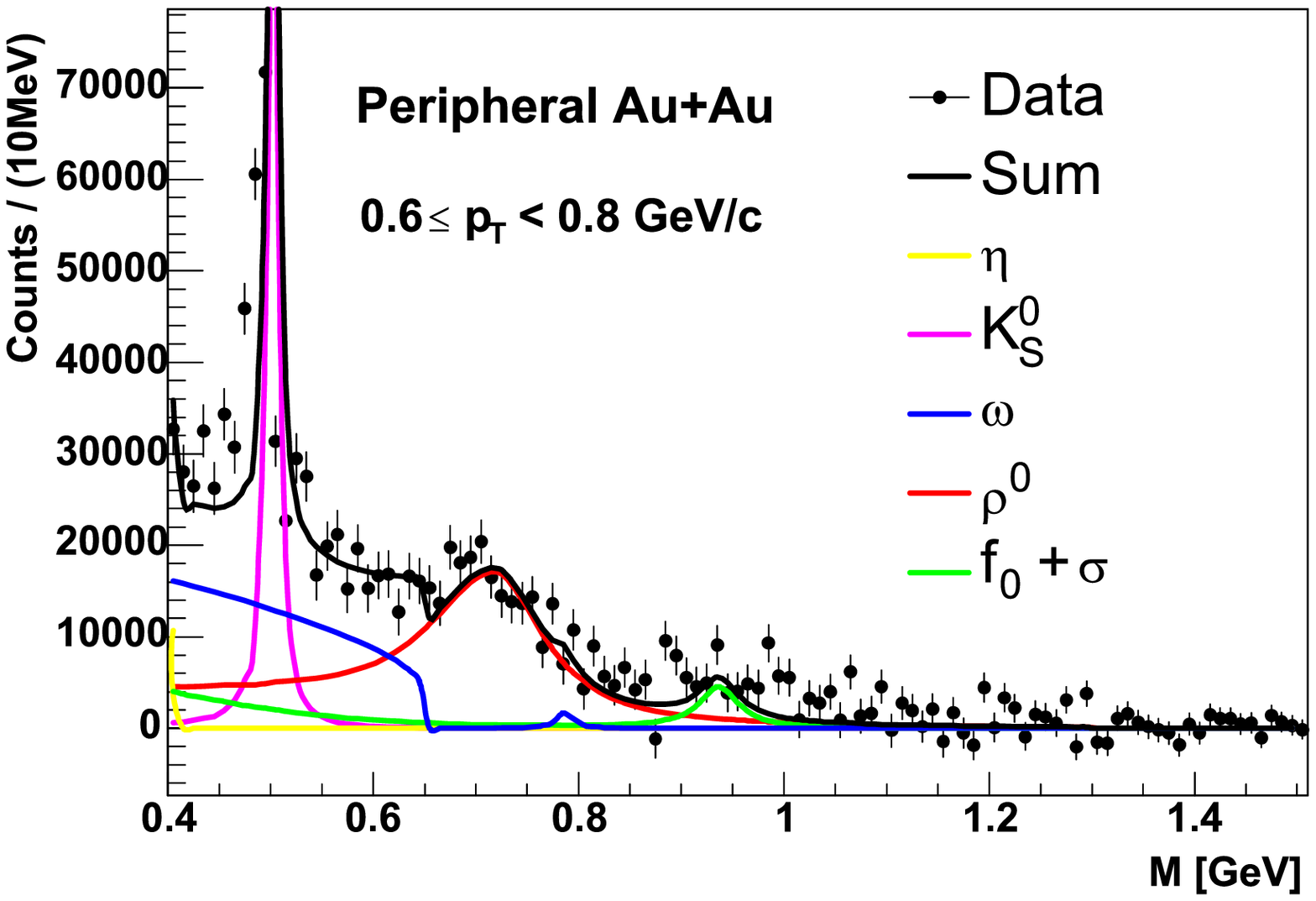,width=5.9cm,height=5.1cm}
\hspace{1cm}
\epsfig{file=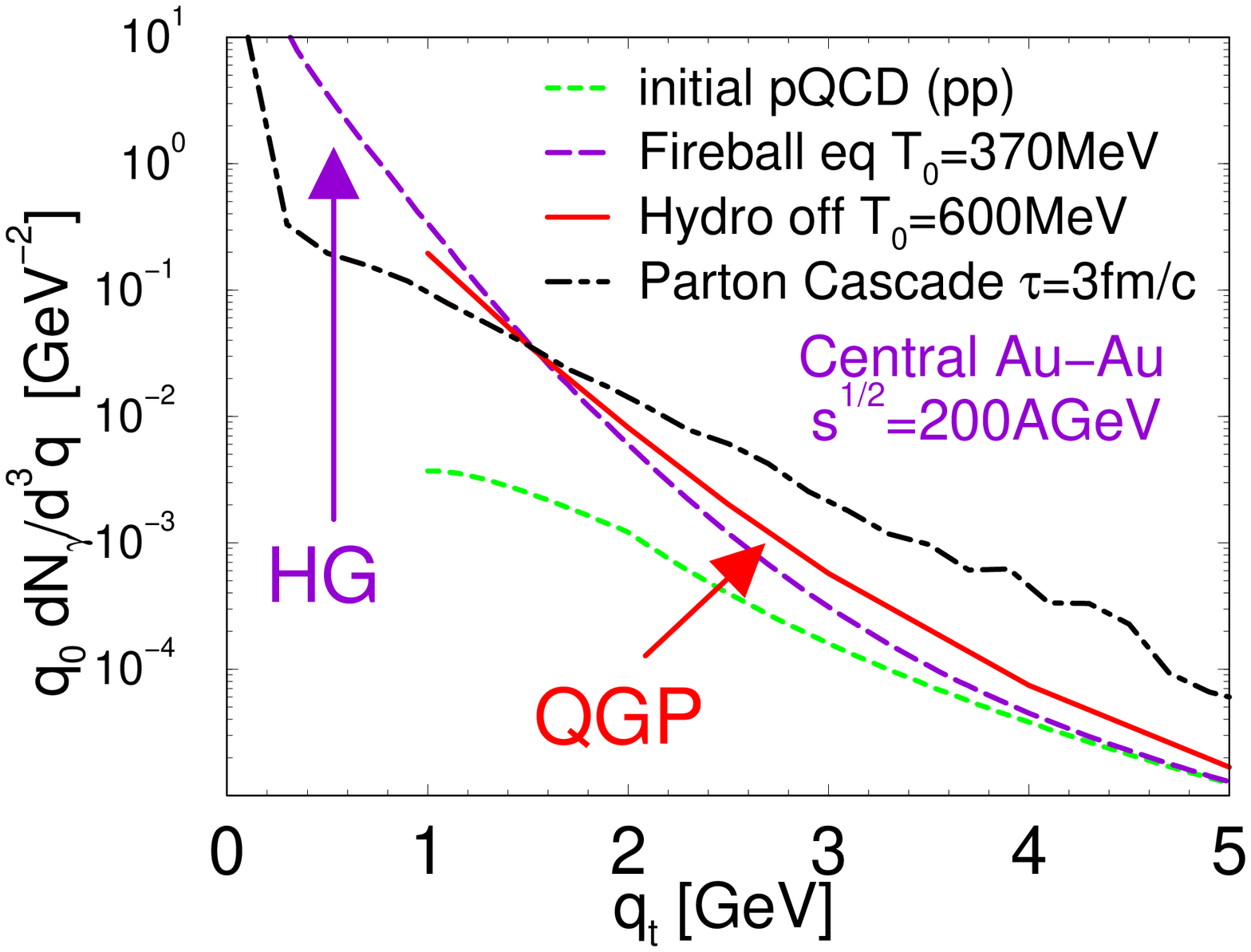,width=6.5cm,height=5.1cm}
\end{center}
\vspace{-0.4cm}
\caption{Left panel: model fit~\cite{BFH03} to STAR~\cite{star-rho} 
$\pi^+$$\pi^-$ invariant-mass spectra with a $\rho$-mass shift of -50~MeV. 
Right panel: direct photon sources at RHIC.}
\label{fig_probes}
\end{figure}
The left panel of Fig.~\ref{fig_probes} shows a model fit to STAR 
data~\cite{star-rho} from 40-80\% central $Au$-$Au$ collisions
employing empirical $\pi\pi$ phase shifts in a sudden breakup scenario
as discussed by Florkowski~\cite{BFH03}. He inferred a $\rho$-mass 
drop of about 50~MeV, similar to previous 
findings~\cite{SB03,KP03}. Broadening effects, Bose-Einstein
correlations and nonresonant background contributions may reduce this 
value significantly~\cite{Ra03,PB03}. Also, a comprehensive understanding
of appreciable mass shifts as extracted from high-multiplicity $p$-$p$ 
or $e^+e^-$ reactions is pending. Preliminary data for the $\Delta^{++}$ 
resonance show a slight increase of its mass, together with a significant 
broadening, when going from peripheral to central 200~AGeV $Au$-$Au$ 
collisions~\cite{star-res}.  

Concerning the resonance yields, measured $\rho/\pi$, $f_0/\pi$ and 
$\Delta/p$ ratios~\cite{star-res} are at least a factor of 2 larger 
than equilibrium values at thermal freezeout~\cite{Ra03}, which could
be suggestive for multiple emission necessitating
a rate equation approach~\cite{SB03,KP03,Ra03}, e.g., 
2-3 generations of emitted $\rho$, $f_0$ and $\Delta$'s. 
Such a scenario can simultaneously account for the observed 
$K^*(892)/K$ ratio~\cite{star-res} -- which is reproduced by 
the equilibrium value at $T_{fo}$~\cite{Ra03} -- since 
$\tau_{K^*}$$\simeq$2-3~$\tau_{\rho,f_0,\Delta}$.     
 
Towards the phase boundary, chiral symmetry restoration 
eventually dictates sub\-stantial medium modifications
in terms of degenerate spectral functions for chiral part\-ners,
e.g., $\pi$-$\sigma$, $\rho$-$a_1$, $N$-$N^*(1535)$. The dilepton
enhancement found at SPS~\cite{ceres04} (to be scrutinized
by NA60~\cite{na60}, as well as PHENIX at RHIC) may well be a related 
signal in the $\rho$$\to$$ee$ channel~\cite{RW00}.   
On the one hand, this poses the challenge to be reconciled with
chemical-freezeout models~\cite{BRS03,BSW03} (usually based 
on vacuum masses), e.g., using self\-consistent thermodynamic 
potentials~\cite{Vosk04}. On the other hand, resonance 
spec\-troscopy may offer 
direct access to pertinent chiral partners, e.g., via 
$a_1$$\to$$\pi\gamma$ or $N^*$$\to$$\eta N$.

\subsection{Electromagnetic (EM) Radiation}
\label{ssec_em}
The penetrating character of EM emission makes it a versatile probe
of all collision stages. It should be emphasized that photons and 
dileptons are (kinematic) facets of the same object, i.e.,
the electromagnetic current correlation function (for $M_\gamma$=0 
and $M_{ee}$$>$0, respectively). 
Gale in his talk advocated that both can be used to tag 
initial energies of jets~\cite{WHS96,SGA03}; the latter also 
radiate electromagnetically in their passage through matter, which,
in particular for $\gamma$'s, could constitute a major source
for $p_t$$\simeq$2-6~GeV~\cite{FMS03}, possibly outshining both pQCD 
and thermal radiation. A characteristic signature of this contribution
is its negative elliptic flow. EM radiation off jets is part of what 
is more generally denoted as ``pre-equilibrium" 
sources, which have recently been evaluated in a parton cascade
simulation as presented in Bass' talk~\cite{BMS03}. The pertinent photon
spec\-tra for RHIC are confronted with hydrodynamic~\cite{Niem04} as 
well as thermal fireball cal\-culations~\cite{TRG04} in 
Fig.~\ref{fig_probes} (right panel), also suggesting that 
pre-equilibrium yields dominate above $p_t$$\simeq$2~GeV. The largest 
component in the parton cascade is due to rescattering of
secondary partons (quantum interference (LPM) effects are not 
included). The two thermal calculations agree rather well, with 
hadron-gas radiation prevalent up to $p_t$$\simeq$1~GeV, where
QGP radiation takes over. It is also noteworthy that che\-mical
off-equilibrium in a QGP affects the thermal photon
yields very little, i.e., undersaturated parton distributions
are essentially compensated by higher temperatures~\cite{Niem04}.
   
WA98 at SPS reported new data on low-$p_t$ (100-300~MeV)
direct pho\-tons~\cite{wa98-lowpt} which exhibit a pronounced excess 
of a factor of 3 or more over thermal fireball calcu\-lations~\cite{TRG04}.
The latter employed a recently improved assessment of hadronic
rates which simultaneously describe CERES low-mass dilepton
data. If confirmed, the low-$p_t$ photon enhancement could point
to appreciable in-medium effects in the late hadronic stages
currently not accounted for, e.g., Bremsstrahlung from $\pi\pi$ 
scattering via in-medium softened ``$\sigma$"-mesons  (a possibly 
related softening might have been observed in low-energy $\pi$-induced 
$\pi$-production off nuclei, $\pi A$$\to$$\pi\pi A$, 
see, e.g., Ref.~\cite{Kuni03}). One should note, however, that
the CERES data at very low mass ($M_{ee}$$\le$100~MeV) do not leave 
much room for strong additional sources~\cite{ceres04}.

\subsection{Charm(onium)}
\label{ssec_charm}
The intermediate scale of the charm-quark mass renders charmed hadrons
another valuable messenger of QGP properties.
The first step is to establish a baseline for their primordial
spectra. A CGC calculation presented by Tuchin~\cite{Tuch03} predicts 
a 40\% decrease in $dN/d\eta$ for open-charm when going from $\eta$=0 
to 2 in central $Au$-$Au$, which is essentially a consequence of the
saturation scale changing from $Q_s$$\le$$m_c$ at $\eta$=0 (implying
$N_{coll}$-scaling) to $Q_s$$\ge$$m_c$ at $\eta$=2 (entailing 
$N_{part}$-scaling). 
On the other hand, in the color-dipole approach of Raufeisen's 
talk~\cite{Rauf03}, a 25\% shadowing is found for midrapidity yields 
which does not significantly vary for $\eta$$\le$2. 

\begin{figure}[!t]
\vspace{-2cm}
\begin{center}
\epsfig{file=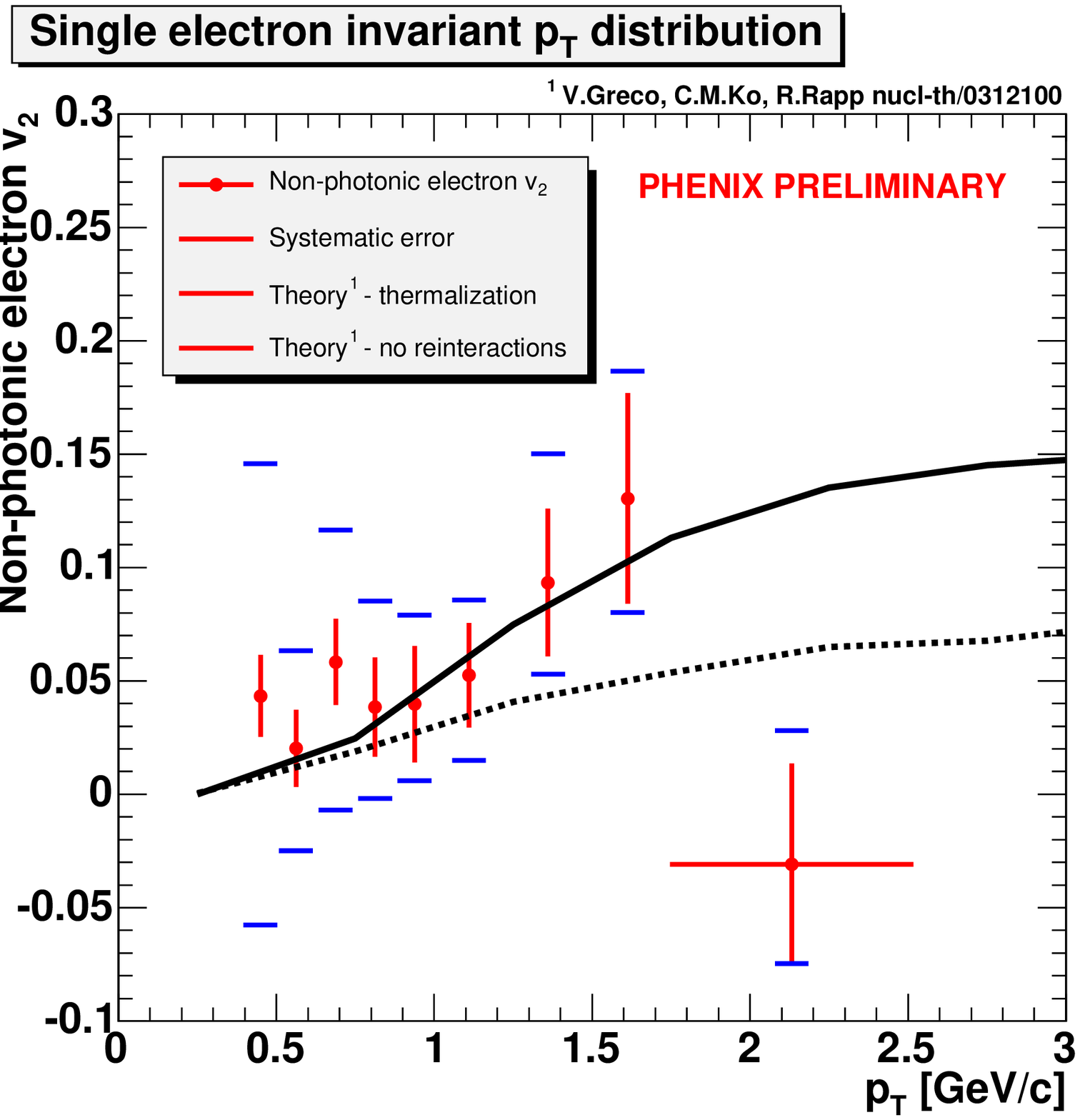,width=6.7cm,height=4.7cm}
\hspace{1cm}
\epsfig{file=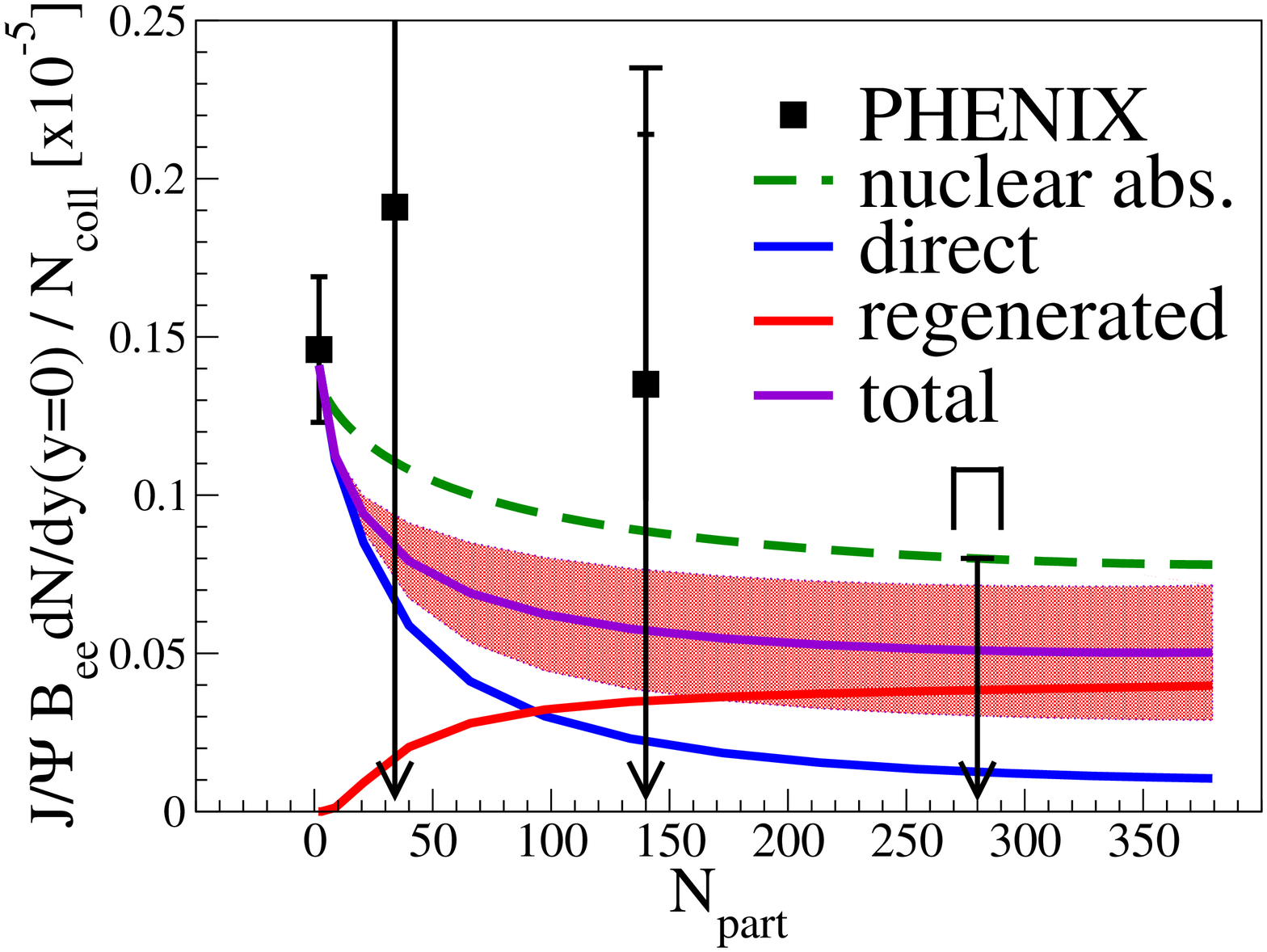,width=7.6cm,height=7.25cm}
\end{center}
\vspace{-0.4cm}
\caption{Left panel: PHENIX elliptic flow data for ``charm-like"
single elec\-trons~\cite{Kelly04} compared to coalescence model
predictions for $D$$\to$$e\nu X$~\cite{GKR03}. Right panel: PHENIX
$J/\psi$ data~\cite{phenix-psi} compared to a kinetic rate equation 
approach~\cite{GRB03}.}
\label{fig_charm}
\end{figure}

Recent PHENIX data for ``non-photonic" single-electron 
$p_t$-spectra~\cite{phenix-e} 
(ascribed to open heavy-flavor decays) in central $Au$-$Au$ are 
consistent with the ``null-effect" on the $c$-quark energy loss, see 
Djordjevic's talk~\cite{Djor03}. However, first PHENIX 
data on the elliptic flow of single-$e^\pm$~\cite{Kelly04}
are suggestive for a non-vanishing signal, cf.~left panel of 
Fig.~\ref{fig_charm}. 
The plot also includes coalescence model predictions~\cite{GKR03} for
parent $D$-mesons formed by recombining thermal light quarks (including 
(elliptic) flow as determined by pion data~\cite{Greco03}) with 
$c$-quarks either from PYTHIA 
(representing no rescattering, dashed line) or being thermalized in, 
and flowing with, the bulk matter (solid line). No definite conclusions 
can be drawn yet. Note that, although the two opposite scenarios of 
no $c$-quark reinteraction vs. thermalization lead to very 
similar {\em single}-$e^\pm$ $p_t$-spectra up to $\sim$2~GeV~\cite{Bats02}, 
pertinent {\em di}-lepton invariant-mass spectra should be quite 
different, since the back-to-back character of hard production
(implying large $M_{ee}$) becomes randomized in the thermal case
(reducing the average $M_{ee}$).
 
Thermalization of $c$-quarks has important impact on charmonium
production. With 10-20 $c\bar c$ pairs in central 200~AGeV $Au$-$Au$, 
and the realization from LQCD that $J/\psi$ states possibly persist
up to 2$T_c$, their regeneration in the QGP becomes plausible, e.g., 
via the inverse of gluon dissociation reactions, 
$J/\psi$+$g\rightleftharpoons c$+$\bar c$+$X$\footnote{This may be 
viewed as a generalization of statistical production at 
$T_c$~\cite{pbm00}, see also Refs.~\cite{Thews01,Zhang02,Cass04}}. 
In particular, thermalized $c$-quarks enable the description of these 
processes in a kinetic theory framework via simplified 
rate equations of type $dN_\psi/dt =\Gamma_\psi (N_\psi^{eq}-N_\psi)$, 
where the reaction rate $\Gamma_\psi$ represents the width of  
in-medium $J/\psi$ spectral functions, and thus, in principle, is
directly amenable to LQCD calculations. In addition, if, as 
to be expected, the open-charm number is conserved in the course 
of an $A$-$A$ reaction (i.e., determined by primordial $N$-$N$ 
collisions), the equilibrium level of charmonia is sensitive to 
open-charm masses~\cite{GRB03}, which can also be determined in LQCD.    
Grandchamp in his talk showed solutions to the kinetic equation
including essential features of LQCD~\cite{GRB03}, cf.~right panel 
of Fig.~\ref{fig_charm}. Most of the $J/\psi$'s in central $Au$-$Au$
at RHIC are indeed regenerated in the QGP, with the band indicating
sensitivities to in-medium open-charm masses.
The observation of a significant number of $J/\psi$'s at RHIC
would be another step towards ascertaining the notion of hadronic 
resonance states in the (nonperturbative) QGP.

\section{Final Remarks}
\label{sec_sum}
The search for the QGP has reached a critical phase: the combination 
of current RHIC data with theoretical analyses shows   
that the matter created in central $Au$-$Au$ collisions is
(i) very dense, (ii) thermalized, and (iii) probably nonperturbative.
Further combining (i)+(ii) with the lattice QCD value for the
critical energy density, one is lead to conclude that the QGP has 
indeed been produced at RHIC. However, to claim such a discovery, 
one would like to have a deeper understanding of the 
nature of that phase, as encoded in (iii), i.e., what the relevant 
(microscopic) degrees of freedom and their interactions are.  
In this respect, upcoming charm(onium) and electromagnetic
probes (photons, dileptons) can be expected to provide
valuable insights.

\vspace{0.5cm}

\noindent{\bf Acknowledgment} \\
I thank the conference organizers for the honor of 
inviting me for the theoretical rap\-porteur talk. 
I am grateful to S.Bass, G.E.Brown, L.Grandchamp, V.Greco, U.Heinz, 
F.Karsch, C.M.Ko, B.Kopeliovich, P.Petreczky, K.Rajagopal, 
K.Redlich, V.Ruuskanen, E.Shu\-ryak, R.Venugopalan and 
I.Vitev for discussion and/or information on their talks.

\end{document}